\documentclass[prd,nofootinbib,reprint,twocolumn]{revtex4-1}
\usepackage{amsmath,amssymb,bm,epsfig,color,graphicx}
\usepackage[mathscr]{eucal}
\usepackage{slashed}
\usepackage{cancel}
\usepackage[colorlinks=true,linkcolor=blue,citecolor=blue,urlcolor=blue]{hyperref}
\usepackage[caption=false]{subfig}
\usepackage{hyperref}
\usepackage{tikz,xcolor}
\usepackage{amsmath}
\usepackage{slashed}
\usepackage{comment}
\usepackage{cancel}
\usepackage{multirow}
\usepackage{slashed}

\renewcommand{\thefootnote}{\fnsymbol{footnote}}

\newcommand{\bea}{\begin{array}}
\newcommand{\eea}{\end{array}}
\newcommand{\beq}{\begin{eqnarray}}
\newcommand{\eeq}{\end{eqnarray}}

\newcommand{\SO}{\textrm{SO}}

\newcommand{\Hsym}{\mathcal{H}}
\newcommand{\Gsym}{\mathcal{G}}

\definecolor{orange}{RGB}{255,100,0}
\definecolor{rosepink}{RGB}{248,100,100}

\begin{document}
\rightline{OU-HET-1190}
\rightline{UT-Komaba/23-5}

\vspace{-0.5cm}

\title{
\vspace{0.5cm}
Cosmological Phase Transitions in Composite Higgs Models
}

\author{
Kohei Fujikura$^{1}$,\footnote{
E-mail address: kfujikura@g.ecc.u-tokyo.ac.jp} 
Yuichiro Nakai$^{2,3}$,\footnote{
E-mail address: ynakai@sjtu.edu.cn} 
Ryosuke Sato$^{4}$\footnote{
E-mail address: rsato@het.phys.sci.osaka-u.ac.jp} 
and Yaoduo Wang$^{2,3}\footnote{
E-mail address: yaoduowang@sjtu.edu.cn}$\\*[10pt]
$^1${\it \normalsize
Graduate School of Arts and Sciences, University of Tokyo, Komaba,\\ Meguro-ku, Tokyo 153-8902,
Japan} \\*[3pt]
$^2${\it \normalsize 
Tsung-Dao Lee Institute, Shanghai Jiao Tong University, \\ 520 Shengrong Road, Shanghai 201210, China} \\*[3pt]
$^3${\it \normalsize 
School of Physics and Astronomy, Shanghai Jiao Tong University, \\ 800 Dongchuan Road, Shanghai 200240, China} \\*[3pt]
$^4${\it \normalsize
Department of Physics, Osaka University, 
Toyonaka, Osaka 560-0043, Japan}  \\*[5pt]
}

\begin{abstract}

We investigate cosmological phase transitions in various composite Higgs models
consisting of four-dimensional asymptotically-free gauge field theories. 
Each model may lead to a confinement-deconfinement transition and a phase transition
associated with the spontaneous breaking of a global symmetry
that realizes the Standard Model Higgs field as a pseudo-Nambu-Goldstone boson.
Based on the argument of universality, we discuss the order of the phase transition
associated with the global symmetry breaking
by studying the renormalization group flow of the corresponding linear sigma model at finite temperature,
which is calculated by utilizing the $\epsilon$-expansion technique at the one-loop order.
Our analysis indicates that some composite Higgs models accommodate
phenomenologically interesting first-order phase transitions.
We also explore the confinement-deconfinement transition in a UV-completed composite Higgs model
based on a $Sp(2N_c)$ gauge theory.
It is found that the first-order phase transition is favored when the number of degrees of freedom for the $Sp(2N_c)$ gauge field is much larger than
that of matter fields in the fundamental representation of $Sp(2N_c)$.
We comment on the gravitational wave signal generated by the confinement-deconfinement transition
and its detectability at future observations.
Our discussions motivate further studies on phase transitions in composite Higgs models
with the use of lattice simulations. 

\end{abstract}

\maketitle

\renewcommand{\thefootnote}{\arabic{footnote}}
\setcounter{footnote}{0}

\section{Introduction}

To unveil the nature of the observed Higgs particle is a key to understand our Universe.
Especially, it is a critical issue to answer the question of
whether the Higgs boson is made of more fundamental constituents or not.
The study of a composite Higgs boson has been initiated by Georgi and Kaplan
\cite{Kaplan:1983fs,Kaplan:1983sm,Georgi:1984af,Dugan:1984hq},
where the Higgs boson is identified as a pseudo-Nambu-Goldstone boson (pNGB)
arising from the spontaneous breaking of a continuous global symmetry
triggered by the dynamics of a new confining gauge field theory,
just like the pion as a pNGB associated with the chiral symmetry breaking in the ordinary QCD
(for reviews, see $e.g.$ refs.~\cite{Contino:2010rs,Bellazzini:2014yua}).
An unbroken subgroup of global symmetry is gauged and identified as the $SU(2)_L \times U(1)_Y$ symmetry
of the Standard Model (SM). 
Thanks to the pNGB nature, the SM-like Higgs boson can be naturally lighter than other composite states.
If the new gauge theory contains only fermions and no scalars,
it can provide a solution to the naturalness problem of the electroweak scale 
and also explain its smallness by dimensional transmutation. 
Various composite Higgs models with different patterns of global symmetry breaking have been proposed so far,
as summarized in table 1 of ref.~\cite{Bellazzini:2014yua},
while the authors of refs.~\cite{Barnard:2013zea,Ferretti:2013kya} discussed
four-dimensional UV descriptions based on purely fermionic gauge theories.

If a composite Higgs model consisting of a new confining gauge field theory is realized in nature,
it may show two (distinct) phase transitions in the early Universe:
(i) a confinement-deconfinement transition and (ii) a phase transition
associated with the spontaneous breaking of a global symmetry.
The ordinary electroweak phase transition may follow from those phase transitions or simultaneously take place.
Although the effect of a new strongly-interacting sector in the model on the electroweak phase transition
has been extensively investigated in literature
\cite{Chala:2016ykx,Chala:2018opy,Bian:2019kmg,Frandsen:2023vhu},
including the application to electroweak baryogenesis~\cite{Konstandin:2011ds,Espinosa:2011eu,Bruggisser:2018mus,Bruggisser:2018mrt,Bruggisser:2022rdm},
the dynamics of the phase transitions (i), (ii) has been largely unexplored in the context of a composite Higgs model,
while its understanding is essential to clarify the evolution of our Universe
and possibly brings about a new application to a cosmological issue
as well as a further prediction of the model that is experimentally tested.\footnote{
If a composite Higgs model respects an approximate conformal symmetry,
the confinement-deconfinement phase transition can be described by the dilaton effective theory \cite{Coradeschi:2013gda,Baratella:2018pxi,Agashe:2019lhy}.
Such a scenario is realized in the five-dimensional dual description of the Randall-Sundrum model
with an appropriate radion stabilization mechanism~\cite{Creminelli:2001th,Randall:2006py,Nardini:2007me,Hassanain:2007js,Konstandin:2011dr,Bunk:2017fic,Konstandin:2010cd,Dillon:2017ctw,vonHarling:2017yew,Fujikura:2019oyi,Agashe:2020lfz}.
Here, we focus on the case where the conformal symmetry is broken by the gauge field theory dynamics,
similar to the case of the ordinary QCD.
}
In particular, if either of the phase transitions (i), (ii) is of the first-order,
it proceeds through the nucleation and expansion of true vacuum bubbles,
which provides a significant departure from the thermal equilibrium,
where the bubble collision~\cite{Turner:1990rc,PhysRevD.46.2384,PhysRevLett.69.2026,Kosowsky:1992vn}, sound wave~\cite{Hindmarsh:2013xza,Giblin:2014qia,Hindmarsh:2015qta,Hindmarsh:2017gnf}
and plasma turbulence~\cite{Kamionkowski:1993fg,Kosowsky:2001xp,Caprini:2006jb,Caprini:2009yp,Gogoberidze:2007an,Niksa:2018ofa}
may generate an observable stochastic gravitational wave (GW) background.
In addition, if there exists a conserved global charge such as a dark baryon number with its finite density,
$i.e.$ a non-vanishing chemical potential, a macroscopic compact object, called a dark quark nugget,
can be formed and become a suitable dark matter candidate~\cite{Bai:2018dxf} (see also ref.~\cite{Heurtier:2019beu,Baker:2019ndr}
for another production mechanism of dark matter utilizing a first-order phase transition).
In a general context, dark confinement and chiral phase transitions in QCD-like theories
have been studied in refs.~\cite{Schwaller:2015tja,Kahara:2012yr,Reichert:2021cvs,Kang:2021epo}.

The strongly-interacting system of a composite Higgs model does not admit the first-principle analytical calculation
to explore the phase transitions (i), (ii).\footnote{The dynamics of a phase transition
in a weakly-interacting system can be directly studied by means of the equilibrium thermal field theory
with the imaginary time formalism~\cite{Dolan:1973qd} 
unless the phase transition is of the second-order or weakly first-order
(see ref.~\cite{Quiros:1999jp} for an excellent review).}
The only possible direct approach is a numerical lattice simulation,
as performed for the electroweak and QCD phase transitions, revealing that they are 
likely to be smooth crossover transitions~\cite{Kajantie:1995kf,Kajantie:1996qd,Rummukainen:1998as,Aoki:2006we}.\footnote {
There remains a possibility of the first-order QCD phase transition
\cite{Yonekura:2019vyz}.}
However, various theoretical approaches have been discussed to clarify the nature of phase transitions
for ordinary QCD and QCD-like gauge theories.
The most famous attempt is to use the argument of universality (see ref.~\cite{Cardy:1996xt}
for a detailed discussion, but we will review it
in the next section).
This approach assumes that long-wavelength fluctuations are dominant during a phase transition,
which is presumably realized when the phase transition is of the second-order or weakly first-order.
Under this assumption, the phase transition may be insensitive to the microscopic physics,
which enables us to determine the order of the phase transition
by studying the effective linear sigma model that respects the symmetries of the system.
The advantage of this approach is that it can be applied to a wide range of systems,
including the ferromagnetic system in condensed matter physics, and the result is largely model-independent
because the analysis only depends on the space dimension as well as the underlying symmetries.
There are attempts based on the argument of universality
for the ordinary QCD~\cite{Pisarski:1983ms,Nakayama:2014sba} and QCD-like theories~\cite{Wirstam:1999ds,Basile:2004wa}.
In the present paper, we discuss the order of the phase transition
associated with the global symmetry breaking in each of various composite Higgs models
by studying the renormalization group (RG) flow of the corresponding linear sigma model at finite temperature,
which is calculated by using the so-called $\epsilon$-expansion technique at the one-loop order.
It turns out that several composite Higgs models favor the first-order phase transitions
within the framework of the argument of universality.

To study a confinement-deconfinement phase transition in a composite Higgs model,
one needs to specify its UV description.
Here, our benchmark model is given by a four-dimensional asymptotically-free $Sp(2N_c)$ gauge field theory presented in ref.~\cite{Barnard:2013zea}.
Unfortunately, it is difficult to analyze the confinement-deconfinement transition
for such a $Sp(2N_c)$ gauge field theory with dynamical matter fields by using a method other than a lattice simulation.
Then, our approach is to take the large $N_c$ limit with a fixed number of flavors and assume that the system is well-described by the pure $Sp(2N_c)$ gauge theory at finite temperature.
In this case, the first-order confinement-deconfinement phase transition is favored from direct lattice calculations~\cite{Holland:2003kg,Braun:2010cy}, and 
the dynamics of the phase transition may be described by
a phenomenological effective theory,
called the Polyakov loop model, that is constructed
in terms of the Polyakov loop as an appropriate order parameter of the phase transition~\cite{Pisarski:2000eq,Pisarski:2001pe}.
By utilizing the result of lattice simulations for the pure $SU(N_c)$ gauge theory,
which is justified at least in the large $N_c$ and zero-temperature limit,
one can quantitatively analyze the first-order confinement-deconfinement phase transition by the Polyakov loop model.
In particular, the use of the Polyakov loop model enables us to derive the gravitational wave spectrum generated by
the first-order confinement-deconfinement transition.
We will discuss its detectability at future observations.

The rest of the present paper is organized as follows.
In section~\ref{sec:the argument of universality},
we determine the order of the phase transition associated with the global symmetry breaking
in each of different composite Higgs models by assuming the argument of universality and
considering the corresponding linear sigma model.
In section~\ref{confinement transition},
we focus on a $Sp(2N_c)$ gauge theory with the large $N_c$ limit and study the confinement-deconfinement transition
by utilizing the Polyakov loop model.
The gravitational wave spectrum generated by the first-order phase transition is calculated.
Section~\ref{sec:conclusion} is devoted to conclusions and discussions.
{In appendix~\ref{appendix:NJL}, we perform the analysis of the phase transition
associated with global symmetry breaking
$U(4) \, (SU(4))\to Sp(4)$ by utilizing the Nambu-Jona-Lasinio model.}

\section{Global symmetry breaking 
}\label{sec:the argument of universality}

To discuss the order of the phase transition associated with the spontaneous breaking of a global symmetry $\Gsym$
to its subgroup $\Hsym$ in a composite Higgs model, we assume the argument of universality.
Despite strong interactions,
the dynamics of the phase transition is then described by the corresponding linear sigma model
whose form is solely determined by the symmetry breaking pattern $\Gsym \to \Hsym$,
and as reviewed here, we can investigate the order of the phase transition
by studying the RG flow of the linear sigma model at finite temperature,
calculated by the $\epsilon$-expansion technique.
Since most of proposed composite Higgs models have global symmetry breaking patterns whose associated phase transitions
have been already explored by using the same technique in literature,
their results will be summarized (see table~\ref{table phase transition}). 
We will also present a new analysis of the order of the phase transition in a composite Higgs model
with global symmetry breaking $SO(N)\to SO(M)\times SO(N-M)$ proposed for $N=9$ {and $M=4$} in ref.~\cite{Chang:2003zn}.

\begin{table*}[t]
\begin{center}
	\begin{tabular}{|c|c|c|c|}
\hline
$\Gsym \to \Hsym$&PT dynamics&Model&Order \\
\hline
\multirow{2}{*}{$SO(N)\to SO(N-1)$}& ~\cite{Brezin:1972fb,Ginsparg:1980ef}  & $N=5$~\cite{Agashe:2004rs} & 2'nd\\
&&$N=9$~\cite{Bertuzzo:2012ya}&2'nd\\
\hline
 $SO(9)\to\SO(5) \times \SO(4)$ &  This work&\cite{Chang:2003zn}&1'st\\
\hline

\multirow{2}{*}{$SU(2N) \, (U(2N))\to Sp(2N)$} & \cite{Wirstam:1999ds}  & $N=2$~\cite{Katz:2005au,Gripaios:2009pe,Galloway:2010bp,Barnard:2013zea}&anomaly\\
& &$N=3$~\cite{Low:2002ws,Katz:2005au}&1'st \\
\hline
$SU(N) \, (U(N))\to SO(N)$ & \cite{Basile:2004wa} &$N=5$~\cite{Arkani-Hamed:2002ikv,Katz:2005au,Vecchi:2013bja} &1'st\\
\hline
	\end{tabular} \\
\caption{The order of a phase transition associated with spontaneous breaking of a global
symmetry. The first column denotes the symmetry breaking pattern, $\Gsym  \to \Hsym$.
The second column gives a reference performing the analysis based on the argument of universality,
while the third column summarizes the corresponding composite Higgs models.
The fourth column shows the order of the phase transition, and
``anomaly'' indicates that the order of the phase transition depends on the restoration of the axial anomaly.
See the main text for detailed discussions.
}
\end{center}
\label{table phase transition}
\end{table*}

\subsection{$SO(N)\to SO(N-1)$}\label{sec:argument of universality}

We begin with the most familiar symmetry breaking pattern, $SO(N)\to SO(N-1)$, which is 
realized for $N=5$ in the minimal composite Higgs model~\cite{Agashe:2004rs}
and for $N=9$ in the composite two Higgs doublet model~\cite{Bertuzzo:2012ya},
to describe the outline of the analysis of the phase transition based on the argument of universality.
The symmetry breaking $SO(N)\to SO(N-1)$ can be described
by introducing an order parameter $\Phi_a$ ($a=1,2,\cdots,N$) in the fundamental representation of $SO(N)$.
We now assume that the phase transition dynamics is dominated by the order parameter and its thermal fluctuation.
In this theory, the temporal direction is compactified with period, $\beta =1/T$, where $T$ is the ambient temperature.
The phase transition dynamics is assumed to be dominated by the long-distance physics
whose length scale is longer than $\beta$.
In this case, the temporal direction can be integrated out, and one obtains the three-dimensional effective action
for the order parameter.\footnote{If the underlying theory is weakly-coupled,
one can perturbatively derive
the three-dimensional thermal effective theory from the original short-distance physics.
See e.g. refs.~\cite{Ginsparg:1980ef,Farakos:1994kx,Huang:1995um,Kajantie:1995dw,Niemi:2018asa,}
for the detailed procedure of dimensional reduction in the equilibrium finite-temperature field theory.}
Under these assumptions, the phase transition dynamics may be described by the following three-dimensional effective Lagrangian:
\begin{align}
\mathcal{L}_E	= & \, \dfrac{1}{2}\partial_i \Phi_a \partial_i \Phi_a + \dfrac{1}{2}m^2(T)\Phi_a\Phi_a+\dfrac{\lambda_3}{4}(\Phi_a \Phi_a)^2  \nonumber \\[1ex]
&+\mathcal{O}(\Phi_a \Phi_a)^3 . \label{eq:Heisenberg model}
\end{align}
Here, $i=1,2,3$ denotes the space index, while we truncate operators of $\mathcal{O}(\Phi_a \Phi_a)^3$.
$m^2(T)$ and $\lambda_3$ represent the temperature dependent mass and coupling whose precise values and expressions generically depend on the short-distance physics.
In a strongly-coupled system, we cannot perturbatively calculate these quantities from the underlying theory,
and hence, we treat $m^2(T)$ and $\lambda_3$ as free parameters.
However, $\lambda_3 > 0$ is required for the stability of the potential.\footnote{One cannot exclude the possibility of $\lambda_3<0$. In this case, one needs to include higher order terms of $\mathcal{O}(\Phi_a\Phi_a)^3$
and the mean field analysis indicates the first-order phase transition.
}
The $SO(N)$ symmetry is spontaneously broken to $SO(N-1)$ at zero temperature, $m^2(T=0)<0$ ($\langle \Phi_a\rangle\neq 0 $), while it is restored at high temperature, $m^2(T)>0$ ($\langle \Phi_a \rangle=0$).

Let us first neglect the fluctuation of the order parameter and discuss the order of the phase transition
in terms of the so-called mean field analysis.
In this case, the phase transition is not of the first order since $\langle \Phi_a \rangle $ continuously vanishes
at the critical temperature $T_C$ defined by $m^2(T_C)=0$ as long as $m^2(T)$ is an analytic function of $T$.
However, near the critical temperature $T=T_C$, the IR fluctuation of $\Phi_a$ is in general non-perturbatively large.
In particular, the coupling constant $\lambda_3$ has mass dimension one in three dimensions,
and the effect of the IR fluctuation scales by some power of the dimensionless ratio, $\lambda_3/m(T)$, using the standard power counting~\cite{Arnold:1992rz,Arnold:1994bp} (for $N\simeq 1$),
where the mass of the order parameter $m(T)$ plays the role of an IR cutoff.
As it comes close to the critical temperature, the effect is obviously unsuppressed,
leading to the IR divergence.
Note that the appearance of the IR divergence is a generic feature of the critical phenomena in three dimensions
as indicated in ref.~\cite{Weinberg:1974hy}.
Therefore, the mean field analysis cannot be justified near the critical temperature,
and one needs a more sophisticated analysis which can deal with the IR fluctuation
to determine the order of the phase transition.

When the second-order phase transition takes place in a simple system such as the Ising model,
it has been experimentally known that the correlation length diverges, and consequently, the system exhibits self-similarity at the critical temperature in the long-distance limit~\cite{Cardy:1996xt}.
Self-similarity at long-distance scales implies that the system experiences the scale invariance at the IR,
which may correspond to the existence of an attractive IR fixed point of coupling constants in the effective theory.
Then, we may argue that if there exists a stable IR fixed point in the effective theory,
the system shows the second-order phase transition at the critical temperature (critical point).
In order to find the presence of an IR fixed point including the effect of the IR fluctuation,
one needs the RG analysis.
Interestingly, as originally found by Wilson~\cite{Wilson:1971vs},
the analysis could successfully reproduce singular behaviors of {thermodynamical} quantities
at the second-order phase transition, which are described in terms of critical exponents.
Following this argument, we here assume that the existence of an attractive IR fixed point corresponds to
the occurrence of the second-order phase transition.
On the other hand, if there is no stable IR fixed point,
one expects that the second-order phase transition does not take place.
Indeed, \textcolor{blue}{if} coupling constants in the effective theory flow to an unstable region of the potential, it is conceivable that the {\it fluctuation-induced} first order phase transition takes place as examined in refs.~\cite{Ginsparg:1980ef,Iacobson:1981jm} for generic scalar models.
Here, the terminology of {fluctuation-induced} is added since the fluctuation drives the first-order phase transition, which cannot be seen in the mean field analysis.
We will encounter this situation in the next subsection.

For $N>2$, the effective Lagrangian of Eq.~\eqref{eq:Heisenberg model} is known as the Heisenberg model
which describes the phase transition of the Heisenberg ferromagnetic system in condensed matter physics.
The RG analysis has been carried out by utilizing the $\epsilon$-expansion technique at the one-loop level in refs.~\cite{Wilson:1971vs,Ginsparg:1980ef}.
In the $\epsilon$-expansion, one calculates loop corrections to $\lambda_3$ in $4-\epsilon$ dimensions instead of directly working in three dimensions.
Using the standard $\overline{\rm MS}$ subtraction, the RG equation at $m^2(T=T_C)=0$ is given by~\cite{Peskin:1995ev}
\begin{align}
    \beta_{\lambda_3} \equiv \mu\dfrac{\partial \lambda_3}{\partial\mu} = -\epsilon \lambda_3 + (N+8)\dfrac{\lambda_3^2}{8\pi^2} \, ,
\end{align}
where $\mu$ is the renormalization scale.
It can be seen that there exists a stable fixed point at $\lambda_3^*=8\pi^2\epsilon/(N+8)$,
which is called the Wilson-Fisher fixed point~\cite{Wilson:1971dc}.
We finally obtain the result in three dimensions by the extrapolation of $\epsilon\to 1$.
Since the attractive IR fixed point exists, the phase transition associated with the symmetry breaking $SO(N)\to SO(N-1)$
is expected to be of the second-order, and its property is characterized by the fixed point.
For this symmetry breaking pattern, the presence of the IR fixed point has been also reported by the analysis of $1/N$ expansion~\cite{Brezin:1972se}.

\subsection{$SU(2N) \, (U(2N))\to Sp(2N)$}\label{sec:Sp(2N)}

We next consider phase transitions in composite Higgs models whose symmetry breaking patterns are given by
$SU(2N)\to Sp(2N)$ with $N=2$~\cite{Katz:2005au,Gripaios:2009pe,Galloway:2010bp,Barnard:2013zea} and $N=3$~\cite{Low:2002ws,Katz:2005au}.
Such a global symmetry breaking is realized in a QCD-like theory with $2N$ flavors of quarks
belonging to the pseudo-real representation under a given gauge group~\cite{Peskin:1980gc}.
For $N=2$, ref.~\cite{Barnard:2013zea} has presented a UV completed composite Higgs model that contains
the top partner and satisfies the requirement of anomaly matching.
In this case, one can discuss the chiral phase transition by using the Nambu-Jona-Lasinio (NJL) model
as well as the argument of universality.
We will describe the discussion of the phase transition based on the NJL model in appendix~\ref{appendix:NJL}.

The spontaneous breaking $SU(2N)\to Sp(2N)$ can be described by
an order parameter which belongs to the second-rank anti-symmetric tensor representation of $SU(2N)$, $\Phi_{ab}=-\Phi_{ba}$ ($a,b=1,2,\cdots,2N$).
This field transforms as $\Phi\to U\Phi U^T$ under the $SU(2N)$, where $U$ denotes a $SU(2N)$ matrix.
If $\Phi_{ab}$ gets a vacuum expectation value (VEV) of the form $\Phi_{ab}\propto J_{ab}$ where $J_{ab}$ is the invariant tensor of $Sp(2N)$, the $SU(2N)$ symmetry is broken to $Sp(2N)$.
As in the case of the previous subsection, one can write down the three-dimensional effective theory of the current system as 
\begin{equation}
	\begin{split}
	\mathcal{L}_E = & \, {\rm Tr}\left( \partial_i \Phi^\dag \partial_i \Phi\right) +m^2(T){\rm Tr}\left(\Phi^\dag\Phi\right)+ \dfrac{u}{4}\left({\rm Tr}\left[\Phi^\dag\Phi\right]\right)^2 \\
 &+\dfrac{v}{4} {\rm Tr}\left[\Phi^\dag\Phi\right]^2 +c(T)\left({\rm Pf} (\Phi)+{\rm h.c.}\right), \label{eq:anti-symmetric}
	\end{split}
\end{equation}
where we have neglected higher-dimensional operators of $\mathcal{O}(\Phi^\dag\Phi)^3$,
and ${\rm Pf}(\Phi)$ denotes pfaffian of $\Phi_{ab}$ leading to the $U(1)$ breaking by the axial anomaly.
If $c\,(T_C)=0$, the flavor symmetry is enhanced to $\Gsym = U(2N)$.
Therefore, the universality class of the system is affected by the (non-)presence of the axial anomaly.
At zero temperature, the VEV $\Phi_{ab}\propto J_{ab}$ is realized for $m^2(T=0)<0$
\cite{Li:1973mq,Elias:1975yd}.
The stability of the potential requires $u>0$ and $u+v/N>0$.

The phase transition has been investigated in ref.~\cite{Wirstam:1999ds}
by using the $\epsilon$-expansion technique at the one-loop order
in the context of a $SU(2)$ gauge theory with $2N$ flavors of quarks belonging to the fundamental representation.
Let us first discuss the case without the axial anomaly, i.e. $c(T_C)=0$,
where the symmetry breaking pattern is $U(2N)\to Sp(2N)$.
The RG equations of the effective Lagrangian~\eqref{eq:anti-symmetric} are given by
\cite{Wirstam:1999ds}
\begin{equation}
\begin{split}
	&\beta_u \equiv \mu \dfrac{\partial u}{\partial \mu}= -\epsilon u+\dfrac{2N^2-N+4}{\pi^2}u^2+\dfrac{4N-2}{\pi^2}uv+\dfrac{3}{2\pi^2}v^2 \, , \\[1ex]
	&\beta_v \equiv \mu \dfrac{\partial v}{\partial \mu}= -\epsilon v +\dfrac{4N-5}{2\pi^2}v^2+\dfrac{6}{\pi^2}uv \, .
\end{split}
\end{equation}
In this case, there is no stable IR fixed point for $N>1$,
and the RG flow drives $v$ into the unstable region.
This instability indicates that the phase transition is of the fluctuation-induced first-order.

We next discuss the case with $c(T_C)\neq 0$,
where the symmetry breaking pattern is $SU(2N)\to Sp(2N)$.
In the effective Lagrangian, the most relevant operators (except for the mass term)
are considered to give the dominant effect on the phase transition dynamics.
For $N>4$, the ${\rm Pf}\,(\Phi)$ term becomes less relevant compared to the $u$ and $v$ operators and thus we can neglect it.
Therefore, the fluctuation-induced first-order phase transition takes place for $N>4$.
For $N =3$, the anomaly term is relevant and behaves as a cubic term of $\Phi$ leading to the first-order phase transition.
On the other hand, the order of the phase transition for $N=2$ strongly depends on the effect of the anomaly.
Since $SU(4)$ and $Sp(4)$ are {locally} isomorphic to $SO(6)$ and $SO(5)$, it is speculated in ref.~\cite{Wirstam:1999ds} that the linear sigma model of Eq.~\eqref{eq:anti-symmetric} falls into the $SO(6)$ universality class
which shows the second-order phase transition as discussed in the previous subsection.
In summary, the phase transition is of the first order for $N>2$,
while it can be of the first-order or the second-order for $N=2$ depending on the effect of the anomaly.

\subsection{$SU(N) \, (U(N))\to SO(N)$}

The littlest Higgs model
\cite{Arkani-Hamed:2002ikv,Katz:2005au,Vecchi:2013bja} shows a symmetry breaking pattern,
$SU(N)\to SO(N)$ with $N=5$.
This breaking pattern is realized in a QCD-like theory with $N$ quarks belonging to the real representation
under a given gauge group
\cite{Peskin:1980gc}.
The symmetry breaking $SU(N)\to SO(N)$ can be described by an order parameter which belongs to
the second-rank symmetric tensor representation of $SU(N)$, $\Phi_{ab}=\Phi_{ba}$ ($a,b=1,2,\cdots,N$).
The three-dimensional effective Lagrangian is given by
\begin{equation}
	\begin{split}
	\mathcal{L}_E = & \, {\rm Tr}\left( \partial_i \Phi^\dag \partial_i \Phi\right) +m^2(T){\rm Tr}\left(\Phi^\dag\Phi\right)+ \dfrac{u}{4}\left({\rm Tr}\left[\Phi^\dag\Phi\right]\right)^2 \\
 &+\dfrac{v}{4} {\rm Tr}\left[\Phi^\dag\Phi\right]^2 +c(T)\left({\rm det} (\Phi)+{\rm h.c.}\right), 
\end{split}
\end{equation}
A diagonal VEV, $\Phi_{ab}\propto \delta_{ab}$, drives $SU(N)\to SO(N)$
(or $U(N)\to SO(N)$ if the t'Hooft operator is absent).
For $N>4$, the determinant operator becomes less relevant compared to the $u$ and $v$ operators.
Since we are mostly interested in the case of $N=5$, we neglect the t'Hooft term in the following analysis.

The RG analysis with the $\epsilon$-expansion technique at the one-loop order leads to
\begin{equation}
\begin{split}
		&\beta_u= -\epsilon u + \dfrac{N^2 + N + 8}{24\pi^2}u^2 + \dfrac{N+1}{6\pi^2} uv + \frac{1}{8\pi^2}v^2 \, ,  \\[1ex]
		&\beta_v= -\epsilon v + \dfrac{1}{2\pi^2} uv + \dfrac{2N+5}{24\pi^2}v^2 \, ,
\end{split}
\end{equation}
which indicate that there is no stable IR fixed point, and the RG flow drives the $u$ and $v$ couplings
into the unstable region.
This signals the fluctuation-induced first order phase transition.
{This result is in agreement with that of ref.~\cite{Basile:2004wa}.}

\begin{figure}
	\centering
	\includegraphics[scale = .5]{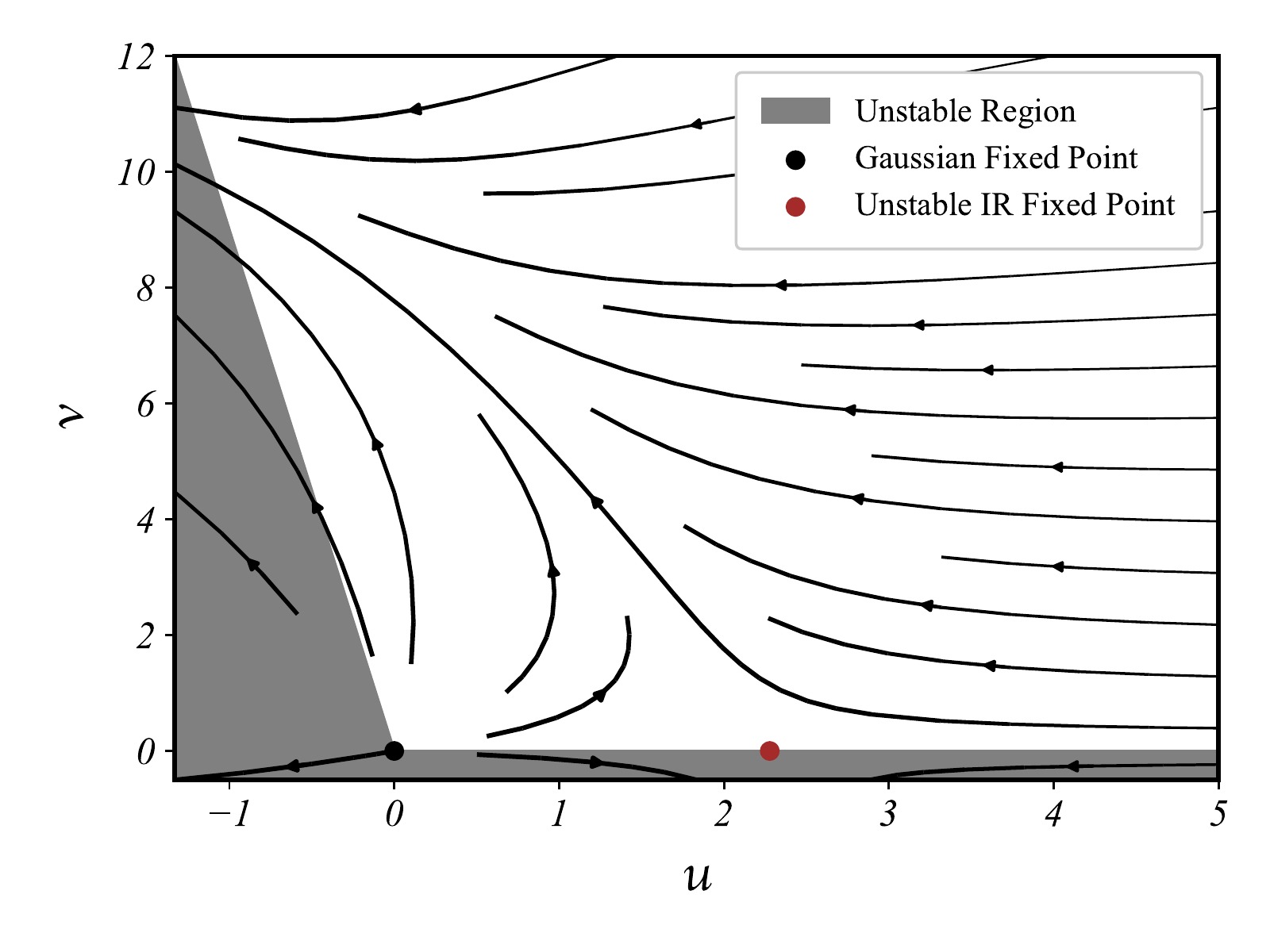}
	\caption{The RG flow of the $u$ and $v$ couplings in the effective Lagrangian~\eqref{eq:symmetric}.
	The potential is not bounded from below in the gray colored region.
	Unstable IR fixed points exist at $(u,v)=(3\pi^2/13,0)$ and $(u,v)=(0,0)$ (Gaussian fixed point).}
	\label{fig:rgFlow}
\end{figure}

\subsection{$SO(N) \to SO(M) \times SO(N-M)$}

A composite Higgs model with global symmetry breaking $SO(N)\to SO(M)\times SO(N-M)$ with $M=\lfloor N/2 \rfloor$
has been proposed for $N=9$ in ref.~\cite{Chang:2003zn}.
Such a symmetry breaking pattern is realized by introducing the real bi-fundamental scalar field $\Phi_{ab}$, which satisfies the symmetric and traceless conditions, $\Phi_{ab}=\Phi_{ba}$ and $\rm Tr \Phi=0$.
The effective Lagrangian is

\begin{equation}
	\begin{split}
	\mathcal{L}_E =  
	& \, \frac12 {\rm Tr}\left( \partial_i \Phi\right)^2 
	+\frac12 m^2(T){\rm Tr}\,(\Phi^2)
	\\
    &+ \dfrac{u}{4!}\left({\rm Tr}\Phi^2\right)^2 
	+\dfrac{v}{4!}{\rm Tr}\,(\Phi^4) \, . \label{eq:symmetric}
	\end{split}
\end{equation}
One may introduce the determinant operator, ${\rm det}\,\Phi$, but it is irrelevant for $N=9$.
Hence we omit this term to analyze the phase transition dynamics.
The potential is bounded from below when $v>0$ and $u+v/N>0$ are satisfied.

Let us now discuss the order of the phase transition, which is new to the best of our knowledge.
By using the $\epsilon$-expansion technique, RG equations for $N>6$ at the one-loop order
are
\begin{equation}
\begin{split}
\beta_u =& -\epsilon u + \frac{1}{16\pi^2} \biggl[ \frac23 \left(N^2+N + 14\right)u^2 \\
&+ \left(\frac83 N + 4 -\frac8N\right) uv 
 + \left(2+\frac{12}{N^2}\right)v^2 \biggr], \\[1ex]
\beta_v =& -\epsilon v + \frac{1}{16\pi^2} \left[ 16uv+\left( \frac43 N + 6 - \frac{24}{N}\right)v^2 \right].
\end{split}\label{eq:effective Lagrangian density SO}
\end{equation}
Figure~\ref{fig:rgFlow} shows RG evolutions of the $u$ and $v$ couplings for the case of $N=9$.
We can see that there exist unstable IR fixed points at $u=3\pi^2/13,~v=0$ and $u=v=0$ (Gaussian fixed point).
The figure indicates that the RG flow drives $u$ and $v$ into the unstable region of the potential,
unless bare couplings are tuned to be the values at unstable fixed points.
Therefore, the symmetry breaking $SO(N)\to SO(M)\times SO(N-M)$ with $N =9,~M=4$ is expected to
accommodate the fluctuation-induced first-order phase transition.

\subsection{Comments on the universality argument}

Our analysis to find the order of the phase transition associated with the spontaneous breaking of a global symmetry
is based on two important assumptions:
(i) the effect of an explicit breaking of the global symmetry on the phase transition dynamics
is negligible, and (ii) the most important excitation is the order parameter during the thermal phase transition.
In reality, it is unclear that both of the assumptions are justified.
Indeed, the explicit breaking is required to give the observed SM-like Higgs boson mass.
Moreover, the $SU(2)_W\times U(1)_Y$ gauge fields always couple to the order parameter.
It is usually challenging to take account of these effects in the universality argument.

Ref.~\cite{Thaler:2005kr} has proposed the use of gauge moose~\cite{Georgi:1985hf} to realize desired global symmetry breaking patterns
in composite Higgs models.
In this construction, a global symmetry breaking is realized by gauging a subgroup of an enlarged global symmetry.
In such a case, the effect of gauge fields is clearly essential, and one may not justify our assumption (i).
We do not consider this construction in the present discussion.

Finally, let us comment on an issue of the $\epsilon$-expansion technique.
We have obtained our results by the extrapolation of $\epsilon\to 1$, which is not fully justified in general.
For example, the inclusion of higher order corrections may change our conclusion.
One of the powerful methods to overcome this difficulty is to utilize the conformal bootstrap technique
\cite{El-Showk:2012cjh}
because it does not rely on the perturbative calculation.
The use of the conformal bootstrap approach is beyond the scope of the present paper
and left for a future study.

\section{Confinement transition}
\label{confinement transition}

We now take a benchmark model proposed in ref.~\cite{Barnard:2013zea}
where the strong dynamics of a $Sp(2N_c)$ gauge theory induces the spontaneous breaking of its flavor symmetry
to realize the pNGB Higgs,
and investigate the confinement-deconfinement phase transition.
By taking the large $N_c$ limit with a fixed number of flavors,
the system is assumed to be well-described by the pure $Sp(2N_c)$ gauge theory at finite temperature.
Then, we discuss that the dynamics of the phase transition can be described by
the Polyakov loop model which is constructed
in terms of the Polyakov loop identified as an order parameter of the phase transition~\cite{Pisarski:2000eq,Pisarski:2001pe}.
By using the result of lattice simulations,
one can quantitatively analyze the phase transition.
We can derive a GW spectrum generated by
the first-order confinement-deconfinement transition
and discuss its discovery prospect.

\subsection{Polyakov loop}

In a pure Yang-Mills theory with a unitary gauge group $G$, such as $SU(N_c)$ and $Sp(2N_c)$,
we can define a gauge invariant operator called Polyakov loop
\cite{Polyakov:1978vu},
\begin{align}
	l_P\equiv \dfrac{1}{{\rm dim}(G)}{\rm Tr}_c \left[\bm{L}_P\right],
\end{align}
where ${\rm Tr_c}$ and ${\rm dim}(G)$ denote the trace taken over the color space
and the number of dimensions of the gauge group $G$ in the fundamental representation, respectively.
For example, ${\rm dim}(Sp(2N_c))=2N_c$ and ${\rm dim}(SU(N_c))=N_c$.
The thermal Wilson line $\bm{L}_P$ is defined as
\begin{align}
	\bm{L}_P \equiv \mathcal{P}\exp\left[\int_0^\beta \mathrm{d}\tau \, T^aA^a_4(\tau, \bm{x})\right]. 
\end{align}
Here, $\bm{x},~\tau,~\beta\equiv 1/T$ are three spatial coordinates, the Euclidean time
and the inverse of the ambient temperature, respectively.
A ${\rm dim}(G)\times {\rm dim}(G)$ matrix $T^a$
is a generator of the color gauge group $G$ in the fundamental representation,
and $A_4^a$ is the Euclidean temporal component of the gauge field.
$\mathcal{P}$ denotes the path-ordering along the temporal direction.
In the current normalization, the Euclidean temporal component of the covariant derivative of a matter field
transformed under the fundamental representation
is given by $D_4 \equiv \partial_4 -A_4$.

The thermal average of the Polyakov loop $l_P$ behaves as $\langle l_P\rangle\propto e^{-\beta\Delta F} $
where $\Delta F$ represents the free energy of an isolated test quark relative to the energy without the quark
(see refs.~\cite{McLerran:1980pk,McLerran:1981pb} for the path-integral derivation of this behavior
in the finite temperature field theory).
Then, if the expectation value of the Polyakov loop vanishes, $\langle l_P \rangle =0$,
the free energy of an isolated test quark costs an infinite energy,
which is identified with the confinement phase.
On the other hand, if $\langle l_P\rangle\neq 0$, the free energy of an isolated test quark is finite,
and hence, the system is identified with the deconfinement phase.
Therefore, $l_P$ can be regarded as a good order parameter for the confinement-deconfinement phase transition.

There exists a global symmetry to distinguish the confinement and deconfinement phases.
In the equilibrium thermal field theory (with the imaginary time formalism),
the gauge field satisfies a periodic boundary condition for the Euclidean time,
$A_\mu(\tau,{\bm x}) =A_\mu (\tau +\beta ,{\bm x})$.
The gauge transformation is
\begin{align}
	A_\mu (\tau,{\bm x})\to  U(\tau,{\bm x})(A_\mu (\tau,{\bm x})+i\partial_\mu)U^\dag(\tau,{\bm x})\,, \label{eq:gauge transformation}
\end{align}
with $U(\tau,{\bm x})  \in G$.
A remarkable feature is that the periodicity of the gauge field does not necessarily result in
the periodicity of $U(\tau,{\bm x})$.
The periodicity of the gauge field only requires that $U(\tau,{\bm x})$ satisfy the following twisted boundary condition~\cite{Svetitsky:1982gs}:
\begin{align}
     U(\tau+\beta ,{\bm x}) =z U(\tau,{\bm x}) \, .\label{eq:center transformation}
\end{align}
Here, $z$ is the center of the group $G$ such that $z$ commutes with every element of $G$.
We can explicitly write is as $z= e^{i2\pi k/N_c}$ ($k=1,2,\cdots,N_c-1$) for $G=SU(N_c)$
and $z=-1$ for $G=Sp(2N_c)$, in addition to the trivial transformation, $z=1$.
The gauge transformation~\eqref{eq:gauge transformation} with the condition~\eqref{eq:center transformation}
is conventionally called the center transformation.
The non-trivial center transformation acts on the Polyakov loop as $l_P\to z \,l_P$ with $z\neq 1$,
which tells that $l_P$ is charged under the center transformation.\footnote{
From a modern perspective of generalized symmetries, the center symmetry can be understood as a one-form symmetry \cite{Gaiotto:2014kfa}.}
Hence, the confinement phase $\langle l_P\rangle =0$ is regarded as the center symmetric phase, while the deconfinement phase $\langle l_P\rangle\neq 0$ is identified as the broken phase.

When we introduce massless dynamical matter fields, the center symmetry is explicitly broken or preserved
depending on their representations under the gauge group $G$.
For example, a scalar (or fermion) field in the fundamental representation $\phi(\tau,{\bm x})$
transforms as $\phi(\tau,{\bm x}) \to U(\tau,{\bm x})\phi(\tau,{\bm x})$.
The boundary condition of $\phi(\tau,{\bm x})$ in the equilibrium finite temperature field theory is given by $\phi (\tau+\beta,{\bm x})=\pm \phi (\tau,{\bm x})$ where $+$ and $-$ correspond to
the cases of the scalar and fermion, respectively.
This is incompatible with the twisted boundary condition~\eqref{eq:center transformation}
so that the center symmetry is explicitly broken.
Thus, when massless dynamical matter fields in the fundamental representation are introduced,
$l_P$ is no longer a good order parameter for the confinement-deconfinement phase transition.
With a sizable effect of explicit breaking, $l_P$ acquires a non-vanishing value at any non-zero temperature
which usually makes the transition a smooth crossover,
where thermodynamical quantities are smooth functions of the temperature, rather than a phase transition.\footnote{
This situation is very similar to the crossover in a ferromagnetic system with
an external magnetic field, as argued in ref.~\cite{Gavai:1985vi}.}
In practice, by performing the path integral with respect to dynamical matter fields,
one obtains a contribution to the Polyakov loop potential which explicitly breaks the center symmetry.
A good example is the ordinary QCD.
As we will discuss later, the $SU(3)$ pure Yang-Mills theory leads to
the first-order confinement-deconfinement transition.
However, the lattice simulation of QCD including (highly improved staggered) light quarks indicates that the confinement-deconfinement transition is of the crossover rather than the first-order phase transition~\cite{Bazavov:2011nk}.
On the other hand, dynamical matter fields in the adjoint representation do not break the center symmetry
because the gauge transformation is the same as that of the gauge field which is compatible with the twisted boundary condition~\eqref{eq:center transformation}.
Thus, in this case, $l_P$ remains a good order parameter for the phase transition.

Since our benchmark composite Higgs model~\cite{Barnard:2013zea}
contains $N_f = 4$ Weyl fermions in the fundamental representation of the gauge group $G = Sp(2N_c)$,
the center symmetry is explicitly broken.
Another source of explicit breaking arises when the model accommodates top partners
by introducing vector-like colored and hypercharged fermions
in the two (or higher) index representation under $Sp(2N_c)$.\footnote{
Fermion matter contents in generic composite Higgs models accommodating top partners are summarized in ref.~\cite{Cacciapaglia:2015vrx} by considering the requirement of the t'Hooft anomaly matching condition.}
Therefore, in the benchmark model, the Polyakov loop $l_P$ is not a good order parameter in general.

\subsection{Large $N_c$ limit}\label{sec:large Nc expansion}

The analysis of the confinement-deconfinement transition in our benchmark composite Higgs model is
generically challenging due to non-perturbative effects.
One of successful ideas to overcome this difficulty is to make the color number $N_c$ large
and perform the large $N_c$ expansion, as originally proposed by t'Hooft~\cite{tHooft:1974pnl}.
The large $N_c$ limit is defined as $N_c \to \infty$ with a fixed t'Hooft coupling $\lambda \equiv g^2N_c$,
where $g$ is the gauge coupling.
At a large $N_c$, t'Hooft showed that Feynman diagrammatic calculations are available,
and the $N_c$ dependence is determined by the topology of a diagram (see e.g. ref.~\cite{Coleman:1985rnk} 
for the basic argument of the large $N_c$ expansion at zero temperature).

Let us now consider $N_f$ flavors of quarks in the fundamental representation of the gauge group,
and take the large $N_c$ limit with a fixed $N_f$.
The contribution to the vacuum energy from those quarks scales as $\mathcal{O}(N_cN_f)$,
while that of the gauge field is given by $\mathcal{O}(N_c^2)$~\cite{Coleman:1985rnk}.
Hence, the quark contribution is sub-dominant and suppressed by a factor of $N_f/N_c$
compared to the gauge field contribution as least in the zero-temperature field theory.
This feature may be preserved for the deconfinement phase even in the finite-temperature field theory
since the number of degrees of freedom is the same as that of the zero-temperature field theory.
Thus, the Polyakov loop potential may be dominated by the gauge field and scale as $\mathcal{O}(N_c^2)$,
while the matter contribution is given by $\mathcal{O}(N_cN_f)$ and negligible for $N_f/N_c \ll1 $.

Since Weyl fermions introduced to accommodate top partners are in the two-index anti-symmetric (or higher) representation under the gauge group $G$ and in the fundamental representation under the ordinary $SU(3)_C$,
their number of degrees of freedom is larger than that of the gauge field.
Thus, such fermions give unsuppressed effects on the center symmetry at a large $N_c$,
and we cannot discuss the dynamics of the confinement-deconfinement phase transition
since there has been no available information from the lattice simulation or first-principle approach.
For this reason, we concentrate on the large $N_c$ analysis of the confinement-deconfinement transition
in a composite Higgs model without top partners.

We have argued that the Polyakov loop potential may be dominated by the gauge field contribution at a large $N_c$.
It enables us to construct the effective field theory of the confinement-deconfinement phase transition
in terms of the Polyakov loop with the input of lattice simulations for a pure Yang-Mills theory.
The first attempt has been made by Pisarski in refs.~\cite{Pisarski:2000eq,Pisarski:2001pe}
for the confinement-deconfinement transition in the $SU(N_c)$ pure Yang-Mills theory
(see also ref.~\cite{Svetitsky:1982gs} for the analysis of confinement-deconfinement phase transitions for general gauge theories based on the argument of universality).
Let us apply this effective approach to the case of $G=Sp(2N_c)$.
The center symmetry is $Z_2$ whose transformation is defined by $l_P\to -l_P$, and hence,
the potential of $l_P$ must be symmetric under the $Z_2$ transformation.
Following ref.~\cite{Pisarski:2000eq}, we postulate the simplest polynomial Polyakov loop potential as
\begin{align}
    \dfrac{V_{\rm pure}(l_P)}{T^4} = -\dfrac{a(T)}{2}l_P^2 + b(T)l_P^4+c(T)l_P^6 + d(T)l_P^8  \, ,
    \label{eq:Polyakov loop potential}
\end{align}
where $a(T),b(T)$ and $c(T)$ denote temperature dependent couplings that are undetermined within the effective theory and requires the input of lattice simulations or the first-principle approach.
At a critical temperature $T=T_C$, when $b(T_C)>0$ and $a(T_C)=0$ are satisfied for an arbitrary $c(T_C)$,
the above potential takes the same form as that of the Ising model, and
the phase transition is of the second order as long as the argument of universality
and the form of the Polyakov loop potential~\eqref{eq:Polyakov loop potential} are valid
(see Sec.~\ref{sec:argument of universality} for the discussion of the argument of universality).
On the other hand, the first-order phase transition is realized when $b(T_C)<0$ and $c(T_C)>0$.\footnote {$a(T_C)=b(T_C)=0$ corresponds to the phase boundary of the first and second-order phase transitions, called tricritical point.}
Hence, the actual order of the phase transition depends on microscopic physics.
In this discussion, the last term of the potential is irrelevant,
but it might be needed to reproduce the result of lattice simulations.

For a small number of colors, there exist lattice simulations for the confinement-deconfinement transition
in the $Sp(2N_c)$ pure Yang-Mills theory~\cite{Holland:2003kg,Braun:2010cy}.
It has been found that the phase transition is of the second-order for $N_c=1$,
while it is of the first order for $N_c =2,3$.
There is also interesting theoretical progress to clarify the order of the confinement-deconfinement phase transition
in the $Sp(2N_c)$ pure Yang-Mills theory.
The functional RG approach reveals that the phase transition is of the first order~\cite{Braun:2010cy} for $N_c=2$.
Ref.~\cite{Anber:2014lba} showed that the first-order confinement-deconfinement phase transition is confirmed for $N_c> 1$
by studying the quantum phase transition in supersymmetric gauge theories
under the conjecture that the thermal confinement-deconfinement transition is smoothly connected
by the quantum phase transition against gluino mass deformation.
The similar behavior has been found in the $SU(N_c)$ pure Yang-Mills theory.
A lattice simulation has reported that the confinement-deconfinement phase transition in the $SU(2)$ pure Yang-Mills theory
is of the second-order, and its universality class is in good agreement with
that of the three-dimensional Ising model~\cite{Fingberg:1992ju}
(note that $SU(2)=Sp(2)$ in our notation).
On the other hand, the phase transition becomes the first order
when the number of colors is sufficiently large, $N_c>2$~\cite{Lucini:2002kp,Lucini:2003zr,Panero:2009tv,Datta:2010sq,Lucini:2012wq}.
Some theoretical approaches also indicate the first-order phase transition~\cite{Aharony:2005bq,Poppitz:2012nz}.

It is conceivable that pure Yang-Mills theories with large color degrees of freedom generally 
lead to the first order phase transitions, and the nature of the phase transitions is adequately captured by
the large $N_c$ expansion, independent of the detail structure of $G$.
In fact, at zero temperature, the large $N_c$ behaviors of pure Yang-Mills theories with $G = SU(N_c), SO(N_c), Sp(2N_c)$
are equivalent in the sense that the expectation value of the Wilson loop is the same, as explicitly demonstrated in ref.~\cite{Lovelace:1982hz}.
Although there is no direct proof of this similarity in the finite-temperature field theory,
we assume that it is maintained at finite temperature.
That is, thermodynamical properties of confinement-deconfinement transitions in pure Yang-Mills theories
are independent of $G$ in the large $N_c$ limit.
We utilize this assumption and the input of lattice simulations for the $SU(N_c)$ pure Yang-Mills theory
to determine the parameters in Eq.~\eqref{eq:Polyakov loop potential} for $G=Sp(2N_c)$.

Let us fit the Polyakov loop potential under the assumption that the potential
for the $Sp(2N_c)$ pure Yang-Mills theory has the same form as that of $SU(N_c)$ at a large $N_c$
except for the center symmetry.\footnote{Since the center symmetries of the $Sp(2N_c)$ and $SU(N_c)$ pure Yang-Mills theories
are $Z_2$ and $Z_{N_c}$, respectively, a replacement $l_P^2\to |l_P|^2$ is required.
However, this replacement does not affect the following analysis because thermodynamical quantities do not change.}
Refs.~\cite{Huang:2020crf,Kang:2021epo} have argued that the result of lattice simulations for the first-order
confinement-deconfinement transition can be phenomenologically described
by the four and six-dimensional Polyakov loop potential in Eq.~\eqref{eq:Polyakov loop potential} at a large $N_c$.
Following ref.~\cite{Huang:2020crf}, the Polyakov loop potential for the $SU(8)$ pure Yang-Mills theory is fitted by
the following parameterization,
\begin{align}
	&a_8(T) =a_{80} + a_{81} \left(\dfrac{T_{\rm con}}{T}\right)+ a_{82} \left(\dfrac{T_{\rm con}}{T}\right)^2 \nonumber\\& \qquad\qquad +a_{83} \left(\dfrac{T_{\rm con}}{T}\right)^3+a_{84} \left(\dfrac{T_{\rm con}}{T}\right)^4,
\end{align}
\begin{align}
	&a_{80}=28.7,~a_{81} = -69.8,~a_{82} = 134,~a_{83} = -180 ,\\[1ex]
	&a_{84}=56.1,~b_8=90.5,~c_8 = 157,~d_8 =-68.9 ,
\end{align}
where $T_{\rm con}$ is the critical temperature of the confinement-deconfinement phase transition at which two free energy minima of $\langle l_P \rangle =0$ and $\langle l_P \rangle \neq 0$ separated by a potential barrier are degenerate.
In order to translate the above fitting into the case of the $Sp(2N_c)$ pure Yang-Mills theory,
it is needed to take account of the change of color degrees of freedom.
Since the number of degrees of freedom for $Sp(2N_c)$ is $N_c(2N_c+1)$ while that of $SU(N_c)$ is $N_c^2-1$,
the Polyakov loop potential for $Sp(2N_c)$ in Eq.~\eqref{eq:Polyakov loop potential}
may be fitted by
\begin{align}
&a(T) =a_{0} + a_{1} \left(\dfrac{T_{\rm con}}{T}\right)+ a_{2} \left(\dfrac{T_{\rm con}}{T}\right)^2 \nonumber\\& \qquad\qquad +a_{3} \left(\dfrac{T_{\rm con}}{T}\right)^3+a_{4} \left(\dfrac{T_{\rm con}}{T}\right)^4 , \\
	&a_i(T) = \dfrac{N_c(2N_c+1)}{63} a_{8i}(T) \quad (i=0,\cdots,4),\\
	&b=\dfrac{N_c(2N_c+1)}{63}b_8,\\
	&c=\dfrac{N_c(2N_c+1)}{63}c_8,~d= \dfrac{N_c(2N_c+1)}{63}d_8.
\end{align}
Here, we have assumed the color dependence of the potential as $V_{\rm pure} \propto N_c^2-1$ for $SU(N_c)$ and $V_{\rm pure}\propto N_c(2N_c+1)$ for $Sp(2N_c)$, which is justified at least for the $SU(N_c)$ pure Yang-Mills theory
with a large $N_c$~\cite{Kang:2021epo}.

The total Polyakov loop potential is schematically decomposed as 
\begin{align}
     V_{P}(l_P,T)=V_{\rm pure}(l_P,T,N_c) +V_{\rm matter}(l_P,T,N_c,N_f) \, ,
\end{align}
where $V_{\rm pure}$ and $V_{\rm matter}$ represent contributions from the gauge field in Eq.~\eqref{eq:Polyakov loop potential} and from dynamical matter fields in the fundamental representation, respectively.
The first term in the potential preserves the center symmetry, while the second term breaks it explicitly.
It is clear that $V_{\rm pure}(l_P,T,N_c)\propto 2N_c^2$ at a large $N_c$.
We assume that the second term is proportional to $N_fN_c$ in the large $N_c$ limit with a fixed $N_f$.

\subsection{Gravitational wave signals}

We now discuss GW signals generated from the cosmological first-order phase transition
associated with the confinement of the $Sp(2N_c)$ gauge theory in the large $N_c$ limit
(see refs.~\cite{Caprini:2015zlo,Caprini:2019egz,Hindmarsh:2020hop} and references therein
for reviews of GW signals generated by the first-order phase transition).
Since new fields introduced in the composite Higgs model possess SM gauge quantum numbers,
we assume that they share the same temperature as that of the SM thermal plasma.
When the cosmic temperature is high enough, the Polyakov loop potential has the center symmetry breaking minimum
at $\langle l_P\rangle\neq 0$ corresponding to the deconfinement phase.
As the temperature cools down due to the cosmic expansion, a metastable local minimum appears at $\langle l_P \rangle =0$.
At $T=T_{\rm con}$, two minima are degenerate and separated by a potential barrier.
For a lower temperature $T< T_{\rm con}$, the center symmetry preserving minimum $\langle l_P\rangle =0$ becomes
energetically favorable.
Bubbles then nucleate at some nucleation temperature $T_n$ when tunneling takes place.

The nucleation temperature can be roughly estimated in the following way.
The tunneling probability per unit time and per unit volume is expressed as
\begin{align}
	\Gamma =\mathcal{A}e^{-S_E}, \label{eq:decay rate}
\end{align}
where $S_E$ is the classical configuration called bounce, and $\mathcal{A}$ is the fluctuation around the bounce configuration.
When the temperature is high enough, $S_E$ is obtained by the $O(3)$-symmetric bounce configuration~\cite{Linde:1980tt},
\begin{align}
	S_E= \dfrac{S_3}{T}, \quad S_3=\int \mathrm{d}\bar{r}\, 4\pi \bar{r}^2\left[\dfrac{1}{2}\left(\dfrac{dl^B_P}{d\bar{r}}\right)^2+V_P(l^B_P,T)\right] \label{eq:bounce action}.
\end{align}
Here, the length scale is normalized by the temperature, $\bar{r}\equiv rT$, where $r$ is the length
in three-dimensional polar coordinates.
In the above expression, $l^B_P$ is the solution of the following differential equation:
\begin{align}
	\dfrac{\mathrm{d}^2 {l}^B_P}{\mathrm{d}\bar{r}^2}+\dfrac{2}{\bar{r}}\dfrac{\mathrm{d}l^B_P}{\mathrm{d}\bar{r}}+\dfrac{\partial V_P(l^B_P,T)}{\partial l^B_P} =0 \, ,
\end{align}
under the boundary conditions,
\begin{align}
	\left.\dfrac{\mathrm{d}l^B_P}{\mathrm{d}\bar{r}}\right|_{\bar{r}=0}=0 \, , \quad l^B_P(\bar{r}\to \infty)=l_{PF} \, .
\end{align}
{Here, $l_{PF}\neq 0$ is the position of the local minimum of $V_P$.}
By the dimensional analysis, we set $\mathcal{A}\sim T^4$.
The Hubble parameter during the radiation domination era is 
\begin{align}	
H^2(T)= \dfrac{\rho_{\rm rad}}{3M_{\rm Pl}^2} \, , \quad \rho_{\rm rad}=\dfrac{\pi^2}{30}g_*(T)T^4  \, ,
\end{align}
where $M_{\rm Pl}\simeq 2.4\times 10^{18}\,{\rm GeV}$ denotes the reduced Planck mass,
and $g_*(T)=  g_{\rm SM*}(T) + g_{\rm new*}(T)$ is the effective number of relativistic species of the thermal plasma
before the confinement-deconfinement transition.
Here, $g_{\rm SM*}\simeq 106.75$~\cite{Husdal:2016haj} is the one for the SM sector,
while $g_{\rm new*}$ is that of a new sector introduced in the composite Higgs model.
In the large $N_c$ limit, $g_{\rm new *} \simeq 4N_c^2+\mathcal{O}(N_c N_f)$, where factor 2 comes from the polarization degrees of freedom of the gauge field.
The nucleation temperature can be roughly estimated by $\Gamma(T_n)=H^4(T_n)$.
Assuming that $S_3(T)/T$ is a monotonic function around $T=T_n$, the condition leads to
\begin{align}
	\left.\dfrac{S_3}{T}\right|_{T=T_n}= 137-2\log\left(\dfrac{g_*(T_n)}{100}\right)-4\log \left(\dfrac{T_n}{1\,{\rm TeV}}\right). \label{eq:nucleation temperature}
\end{align}
As the right hand side only depends on the cosmic temperature logarithmically, we find $T_n\simeq T_{\rm con}$.
The critical temperature $T_{\rm con}$ is around the confinement scale which is set to be $T_{\rm con}=1\,{\rm TeV}$.
The condition is less sensitive to the precise values of $T_{\rm con}$ and $g_*$ due to the logarithmic dependence.

In order to compute GW signals, one needs to estimate the amount of the released energy of the first-order phase transition transferred into the bulk kinetic energy of the fluid and the mean separation of bubbles.
The ratio of the amount of the released energy to the energy density of the fluid is parameterized as
\begin{align}
	\alpha \equiv \left[ \dfrac{\Delta V_P-\dfrac{1}{4}T\dfrac{\partial \Delta V_P}{\partial T} }{\rho_{\rm rad}}\right]_{T=T_n}.
\end{align}
{Here, $\Delta V_P \equiv V_P(l_{PF})-V_P(l_{Pt})$ with $l_{Pt}\equiv l_P^B(\bar{r}\to 0)$ being the tunneling point obtained by solving the bounce equation.}
The mean bubble separation normalized by the Hubble parameter is
roughly estimated by the rate of the bubble nucleation probability,
\begin{align}
	\widetilde{\beta}\equiv -\dfrac{1}{H(t_n)}\dfrac{\mathrm{d}}{\mathrm{d}t}\left(\dfrac{S_3}{T} \right)_{t=t_n},
 \label{betatilde}
\end{align}
where $t$ is the cosmic time and $t_n$ is the time when bubbles are nucleated.
Using the relation $\mathrm{d}T/\mathrm{d}t=-TH(T)$, Eq.~\eqref{betatilde} can be rewritten as
\begin{align}
	\widetilde{\beta} = \left.T\dfrac{\mathrm{d}}{\mathrm{d}T}\left(\dfrac{S_3}{T}\right)\right|_{T=T_n}.
\end{align}
The other important quantity is the terminal bubble wall velocity $v_w$.
In weakly-coupled theories, the terminal bubble wall velocity can be estimated by computing the friction from the thermal plasma~\cite{Bodeker:2009qy,Bodeker:2017cim,Hoche:2020ysm,Gouttenoire:2021kjv}.
On the other hand, in strongly-coupled theories, it is challenging to explicitly compute the friction.
For this reason, we here simply assume that Jouguet detonation bubbles, where $v_w>c_s=1/\sqrt{3}$ with $c_s$ being the sound speed, are realized.
Then the bubble wall velocity is determined by the following relation~\cite{Espinosa:2010hh}:
\begin{align}
	v_w= \dfrac{\sqrt{2\alpha/3+\alpha^2}+\sqrt{1/3}}{1+\alpha}.
\end{align}
Note that if the actual wall velocity is slower than $c_s$, the resultant GW signals are suppressed.

The total amplitude of GW signals can be schematically decomposed as 
\begin{align}
	\Omega_{\rm GW} = \Omega_{\rm coll}+\Omega_{\rm sound}+\Omega_{\rm tur},
\end{align}
where $\Omega_{\rm coll},~\Omega_{\rm sound}$ and $\Omega_{\rm tur}$ are contributions from bubble collisions, the sound wave and turbulence of the thermal plasma, respectively.
In our analysis, we simply assume that most of the released energy is converted into the hot thermal plasma by frictions acting on the wall.
Then, the dominant contribution comes from sound waves or turbulence of the thermal plasma.
In this case, numerical calculations~\cite{Caprini:2015zlo} reveal that the contribution of sound waves is considerably larger than that of turbulence.
Hence, we focus on the contribution of sound waves in our analysis.

The contribution to $\Omega_{\rm sound}$ is estimated by numerical calculations~\cite{Hindmarsh:2015qta} and given by 
\begin{align}
&\Omega_{\rm sound}h^2 =\Omega_{\rm peak}h^2\left(\dfrac{f}{f_{\rm sound}}\right)^3 \left(\dfrac{7}{4+3(f/f_{\rm sound})^2}\right)^\frac{7}{2},\nonumber\\
&\Omega_{\rm peak}h^2= 2.65\times 10^{-16} \left(\dfrac{10^5}{\widetilde{\beta}}\right)\left(\dfrac{\kappa_{\rm sound}^2}{10^{-5}}\right)\left(\dfrac{\alpha}{1+\alpha}\right)^2\left(\frac{100}{g_*}\right)^{\frac{1}{3}},\\
&f_{\rm sound} = 1.9\times 10^{1}\,{\rm Hz}\left(\dfrac{\widetilde{\beta}}{10^5}\right)\left(\dfrac{1}{v_w}\right)\left(\dfrac{T_n}{1\,{\rm TeV}}\right)\left(\dfrac{g_*}{100}\right)^\frac{1}{6}.\nonumber
\end{align}
Here, $h,~\kappa_{\rm sound}$ and $f$ are the dimensionless Hubble parameter at present time, the fraction of the released energy injected into the energy of GW signals and the frequency of GW signals, respectively.
The fraction $\kappa_{\rm sound}$ can be further decomposed as 
\begin{align}
     \kappa_{\rm sound} = \sqrt{\tau_{\rm sound}} \, \kappa \, ,
\end{align}
where $\tau_{\rm sound}$ and $\kappa$ are the sound-wave period normalized by the inverse of the Hubble parameter and the efficiency coefficient, respectively.
Refs.~\cite{Ellis:2018mja} have pointed out that a suppression factor arises if the sound-wave period is shorter than the inverse of the Hubble parameter during the phase transition.
The sound-wave period $\tau_{\rm sound}$ is expressed as~\cite{Ellis:2018mja,Ellis:2019oqb,Ellis:2020awk}
(see also \cite{Guo:2020grp})
\begin{align}
	\tau_{\rm sound} =
 {\rm min}\left\{1,\dfrac{(8\pi)^{\frac{1}{3}}{\rm Max}\{v_w,c_s\}}{\widetilde{\beta}\bar{U}_f}\right\} \, ,
\end{align}
with $\bar{U}_f$ being the root-mean-square four velocity of the thermal plasma~\cite{Hindmarsh:2015qta} which is approximately given by
\begin{align}
	\bar{U}_f^2 \simeq \dfrac{3}{4}\dfrac{\alpha}{1+\alpha}\kappa\,.
\end{align}
For the Jouguet detonation bubble, the efficiency coefficient can be fitted by~\cite{Espinosa:2010hh}
\begin{align}
	\kappa= \dfrac{\sqrt{\alpha}}{0.135+\sqrt{0.98+\alpha}} \, .
\end{align}

Let us discuss the detectability of GW signals generated by the first-order confinement-deconfinement phase transition
in our benchmark composite Higgs model at a large $N_c$.
By calculating the bounce action~\eqref{eq:bounce action} and
evaluating the nucleation temperature~\eqref{eq:nucleation temperature},
we find $\alpha$ and $\widetilde{\beta}$.
Since the latent heat is proportional to $4N_c^2$ for the $Sp(2N_c)$ pure Yang-Mills theory,
$\alpha= \mathcal{O}(0.1)$ is realized.
It turns out that the duration of the phase transition is maximized at $N_c=9$ which leads to $\widetilde{\beta}=4.3\times10^4$.
With this optimized parameter set, the peak amplitude of GW signals is $\Omega_{\rm peak}h^2\simeq 1.4\times 10^{-15}$
which receives a strong suppression due to a large $\widetilde{\beta}$,
while the peak frequency is $f_{\rm sound}\sim 10\,\mathrm{Hz}$.
Unfortunately, such GW signals are too weak to be detected by future-planned experiments.

\section{Discussion}\label{sec:conclusion}

In the present paper, we have discussed cosmological phase transitions in various composite Higgs models,
each of which may show a confinement-deconfinement transition and a phase transition
associated with the spontaneous breaking of a global symmetry that realizes the SM Higgs field as a pNGB. 
To determine the order of the phase transition for a global symmetry breaking,
we have assumed the argument of universality and studied the effective linear sigma model.
The effect of infrared fluctuations on the phase transition dynamics was taken into account
by the RG analysis with $\epsilon$-expansion at the one-loop order.
For a confinement-deconfinement phase transition, we took the UV-completed model
proposed in ref.~\cite{Barnard:2013zea} as a benchmark.
The model consists of a strongly-coupled $Sp(2N_c)$ gauge theory.
Although the presence of dynamical matter fields in the fundamental representation
makes it difficult to investigate the phase transition,
taking a large number of color degrees of freedom with a fixed number of flavors,
we have argued that the effect of dynamical matter fields is subdominant,
and the first-order confinement-deconfinement phase transition takes place,
as it is favored in the $Sp(2N_c)$ pure Yang-Mills theory for $N_c>1$.
The amplitude of GW signals generated by the first-order phase transition is not
within the reach of future-planned experiments.

So far, we have separately discussed the phase transition associated with a global symmetry breaking
and the confinement-deconfinement transition.
If the size of a gauge group is sufficiently small, it is possible to construct the effective theory which simultaneously describes those phase transitions in terms of the Polyakov loop and the quark condensate,
assuming that the global symmetry breaking is realized by the NJL mechanism.
This Polyakov-Nambu-Jona-Lasinio (PNJL) model has been applied to the chiral
and confinement phase transitions in the ordinary QCD~\cite{Fukushima:2003fw} and QCD-like theories~\cite{Reichert:2021cvs} (see also ref.~\cite{Fukushima:2017csk} for an excellent review).
The usage of the PNJL model is promising to simultaneously analyze both phase transitions
in a UV-completed composite Higgs model.
One cannot apply this approach with a single Polyakov loop $l_P$ alone in contrast to the Polyakov loop model
when the number of color degrees of freedom is large, as argued in ref.~\cite{Reichert:2021cvs}.
Then, the analysis of two phase transitions requires an extension of the PNJL model
such as the matrix model approach~\cite{Halverson:2020xpg,Meisinger:2001cq,Dumitru:2010mj,Dumitru:2012fw}.

\section*{Acknowledgement}

We would like to thank Hiromasa Watanabe for valuable discussions.
KF is supported
by JSPS Grant-in-Aid for Research Fellows Grant No. 22J00345. 
YN is supported by Natural Science Foundation of China No.~12150610465.
The work of RS is supported in part by JSPS KAKENHI No.~23K03415.
\appendix

\section{NJL analysis of $SU(4) \, (U(4)) \to Sp(4)$}\label{appendix:NJL}

We here consider a UV completed model proposed in ref.~\cite{Barnard:2013zea} and
discuss the phase transition associated with global symmetry breaking, $SU(4) \, (U(4))\to Sp(4)$.
As discussed in ref.~\cite{Barnard:2013zea}, the phase transition is induced by four-Fermi interactions.
We extend their analysis by including the effect of thermal fluctuations of new quarks
to determine the order of the phase transition.

The gauge group is $G=Sp(2N_c)$, and the model contains four Weyl fermions $Q_i$ ($i=1,\cdots,4$)
in the fundamental representation.
Since the number of flavors is even and the fundamental representation of $Sp(2N_c)$ is pseudo-real,
the theory does not suffer from both Witten and chiral anomalies.
We introduce the following four-Fermi interactions of the NJL model:
\begin{align}
    \mathcal{L}_{Q} =  & \, \frac{G_1}{2N_c}\left(Q_{i}^aQ_{ja}\bar{Q}^b_{j}\bar{Q}_{ib} \right) \nonumber  \\
    &+ \dfrac{G_2}{8N_c}\left(Q^a_{i}Q_{ja}Q^b_{k}Q_{lb}\epsilon^{ijkl}+{\rm h.c.}\right) . \label{eq:four-fermi interaction}
\end{align}
Here, $\epsilon^{ijkl}$ is the Levi-Civita symbol, and $\bar{Q}$ denotes the complex conjugate of $Q$.
The contraction with respect to $a,b$ is taken by the $Sp(2N_c)$ invariant metric tensor $J_{ab}$ whose matrix form is defined by 
\begin{align}
	J=
	\begin{pmatrix}
		0&{\mathbf 1}_{N_c\times N_c}\\
		-{\mathbf 1}_{N_c\times N_c}&0
	\end{pmatrix},
\end{align}
where ${\mathbf 1}_{N_c\times N_c}$ denotes the $N_c\times N_c$ unit matrix.
The interactions~\eqref{eq:four-fermi interaction} possess the $SU(4)$ flavor symmetry
whose transformation is
defined as $Q_{i}\to U^{(4)}_{ij}Q_{j}$ where $U^{(4)}_{ij}$ is a $SU(4)$ matrix.
For $G_2 =0$, the flavor symmetry is enhanced to $U(4)$.

\begin{figure}[t]
\centering\includegraphics[width=8.5cm]{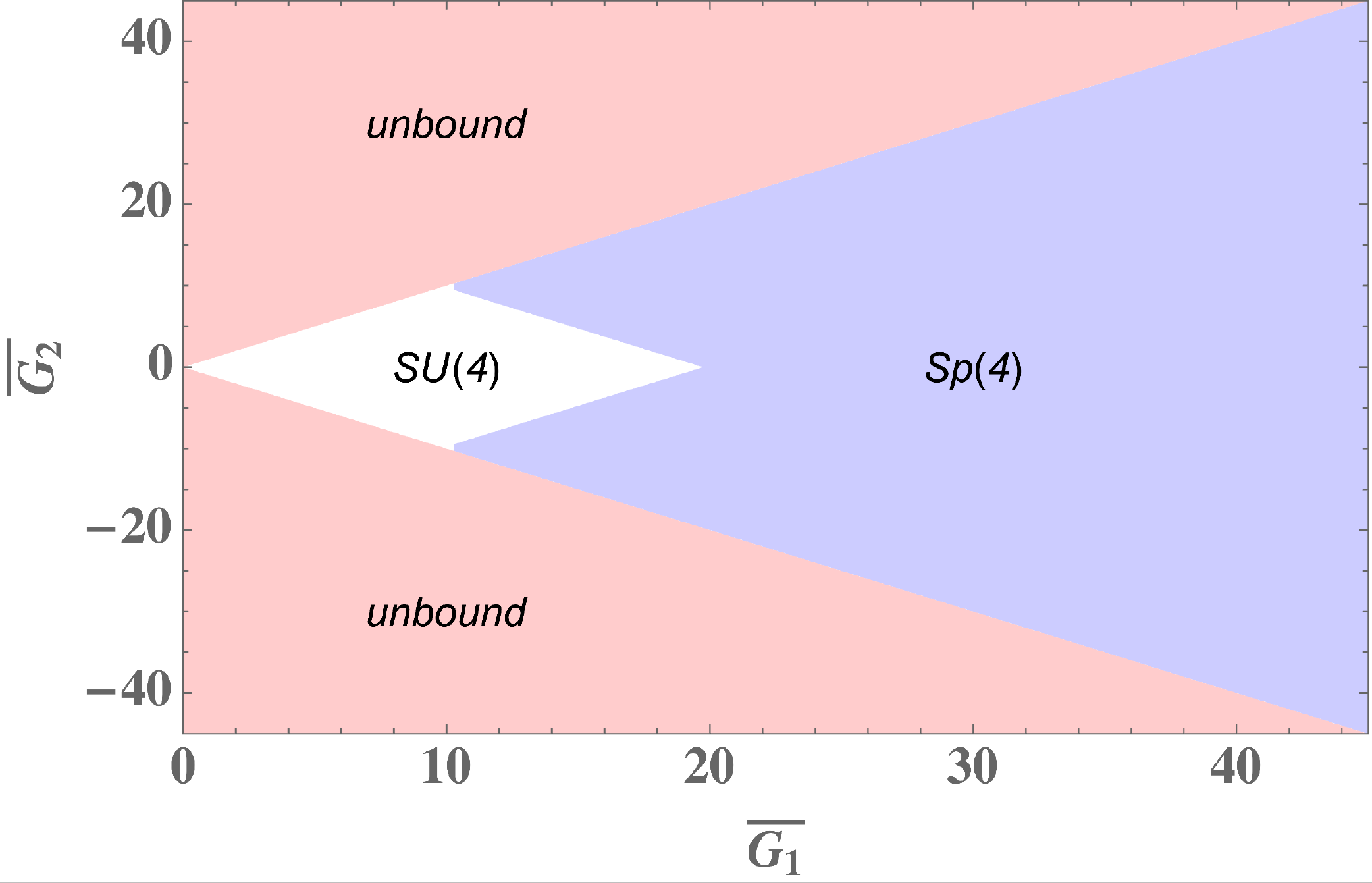}
\caption{
The phase diagram in terms of $\bar{G}_1$ and $\bar{G}_2$. Symmetries of the ground state are $SU(4)$ and $Sp(4)$ for the white and blue colored regions, respectively. In the red colored region, the zero-temperature effective potential is not bounded from below.
}\label{fig:field configurations of two local strings}
\end{figure}

We evaluate the mean-field values of chiral condensate parameterized by auxiliary fields
$M_{ij}\equiv \langle Q_{i}^aQ_{ja}\rangle/N_c$.
At the tree level, the potential for $M_{ij}$ is
\begin{align}
	V^{\rm NJL}_{\rm tree}(M_{ij}) = & \, \dfrac{G_1N_c}{2} {\rm Tr}\left[M M^\dag\right] \nonumber  \\
 &+\dfrac{G_2N_c}{8}\left(\epsilon^{ijkl}M_{ij}M_{kl} +{\rm h.c.}\right).
\end{align}
Following the original analysis of the NJL model, we decompose $Q^a_{i}Q_{ja}$ into mean-field values and
fluctuations, $Q^a_{i}Q_{ja} \to M_{ij}+Q^a_{i}Q_{ja}$.
Then, the effective interactions of $Q$ with $M$ are described by
\begin{align}
	\mathcal{L}_{\rm int} = &- V^{\rm NJL}_{\rm tree} + \dfrac{G_1}{2}\left\{{\rm Tr}\left[Q^aQ_{a} M^\dag\right] +{\rm h.c.}\right\} \nonumber \\
 &+ \dfrac{G_2}{4}\left(\epsilon^{ijkl}Q_{i}Q_{j}M_{kl} +{\rm h.c.}\right).
\end{align}
Since $M_{ij}$ is a $4\times 4$ complex anti-symmetric matrix, using the $SU(4)$ symmetry,
one can parametrize the matrix as
\begin{align}
	M = \begin{pmatrix}
 	0&M_1&0&0\\
 	-M_1&0&0&0\\
 	0&0&0&M_2\\
 	0&0&-M_2&0
 \end{pmatrix},
\end{align}
where $M_1$ and $M_2$ are generally complex.
By integrating out the $Q$ fields,
one obtains the zero-temperature effective potential up to the one-loop order,
\begin{equation}
\begin{split}
    &V^{\rm NJL}_{\rm zero} = V^{\rm NJL}_{\rm tree}+V^{\rm NJL}_{\rm one-loop} \, , \\[1ex]
    &V^{\rm NJL}_{\rm tree}= N_cG_1\left(|M_1|^2+|M_2|^2\right) + N_c G_2(M_1M_2+{\rm h.c.})\,,\\[1ex]
    &V^{\rm NJL}_{\rm one-loop} = -4N_c \int \dfrac{\mathrm{d}^3 k}{(2\pi)^3}\sum_{i=1,2}\sqrt{k^2+|m_i|^2} \, ,\label{eq:NJL effective potential without regularization}\\[1ex]
    &m_1 \equiv e^{i\phi_1}\left|G_1M^*_1+G_2M_2\right|,\\
    &m_2 \equiv e^{i\phi_2}\left|G_1M^*_2+G_2M_1\right| \, .
\end{split}
\end{equation}
Here, $e^{i\phi_{1,2}}$ are arbitrary phases.
The one-loop effective potential is UV divergent and requires renormalization.
Following ref.~\cite{Reichert:2021cvs}, we regularize the UV divergence
by inserting a sharp three-dimensional momentum cut-off $\Lambda_{\rm 3D}$.
Such a regularization scheme is different from that of ref.~\cite{Barnard:2013zea}
where a four-dimensional momentum cutoff is introduced.
With our regularization scheme, the one-loop effective potential
in Eq.~\eqref{eq:NJL effective potential without regularization} is evaluated as
\begin{align}
	V^{\rm NJL}_{\rm one-loop} = &-\sum_{i=1,2}\dfrac{N_c\Lambda^4_{\rm 3D}}{4\pi^2}\biggl[(2+\xi_i^2)\sqrt{1+\xi_i^2} \nonumber \\
 &+\dfrac{\xi_i^4}{2}\log\left(\dfrac{\sqrt{1+\xi_i^2}-1}{\sqrt{1+\xi_i^2}+1}\right)\biggr],
 \quad \xi_i^2\equiv\dfrac{|m_i|^2}{\Lambda^2_{\rm 3D}}\,. \label{eq:one-loop potential NJL}
\end{align}
It is useful to rewrite $V_{\rm tree}^{\rm NJL}$ in terms of $m_{1,2}$,
\begin{align}
    V^{\rm NJL}_{\rm tree}=	\dfrac{N_c}{G_1^2-G_2^2}\left[G_1\left(|m_1|^2+|m_2|^2\right)-G_2\left(m_1m_2+{\rm h.c.}\right)\right]. \label{eq:tree NJL model}
\end{align}
Since $m_{1}$ and $m_2$ are complex variables, $V_{\rm zero}^{\rm NJL}$ depends on three real fields,
$|m_{1,2}|$ and a relative phase of $m_{1,2}$.

A potential for the relative phase comes from the second term of Eq.~\eqref{eq:tree NJL model},
and is minimized for $m_1m_2=\pm|m_1||m_2|$ for $G_2/(G_1^2-G_2^2)\gtrless0$, respectively.
Focusing on the minimum of the relative phase of $m_{1,2}$,
$V^{\rm NJL}_{\rm tree}$ can be reexpressed as 
\begin{align}
	 V^{\rm NJL}_{\rm tree}=	\dfrac{N_cG_1\Lambda^2_{\rm 3D}}{G_1^2-G_2^2}\left(\xi_1^2+\xi_2^2\right)-\left|\dfrac{2N_cG_2\Lambda_{\rm 3D}^2}{G_1^2-G_2^2}\right|\xi_1\xi_2\,.\label{eq:tree-level potential NJL}
\end{align}
Therefore, we effectively need to consider two fields $|m_{1,2}|$ in our analysis.
This discussion is applicable when we include the thermal effect
since the thermal potential of $m_{1,2}$ only depends on $|m_{1,2}|$
(see the concrete expression~\eqref{eq:thermal potential NJL}).
To investigate the parameter region where the chiral symmetry breaking, $\xi_1=\xi_2\neq 0$, is realized,
we numerically evaluate the zero-temperature potential, $\bar{V}^{\rm NJL}_{\rm zero}(\bar{G}_1,\bar{G}_2,\xi_1,\xi_2)\equiv V^{\rm NJL}_{\rm zero} / N_c\Lambda_{\rm 3D}^4$
with $\bar{G}_1\equiv G_1\Lambda_{\rm 3D}^2$ and $\bar{G}_2\equiv G_2 \Lambda^2_{\rm 3D}$.
Fig.~\ref{fig:field configurations of two local strings} displays the phase diagram in terms of $\bar{G}_1$ and $\bar{G}_2$.
We can see from the figure that the chiral symmetry breaking, $SU(4)\to Sp(4)$, takes place for sufficiently large four-Fermi interactions,
while it does not for small interactions.
In fact, the chiral symmetry breaking takes place when the following conditions are satisfied:
\begin{align}
      \bar{G}_1>|\bar{G}_2|\,, \quad 2\pi^2\left(\dfrac{\bar{G}_1-|\bar{G}_2|}{\bar{G}_1^2-\bar{G}_2^2}\right)< 1\,. \label{eq:NJL condition}
\end{align}
The first condition is required for the stability.
We focus on the parameter region to satisfy the conditions~\eqref{eq:NJL condition} in the following analysis.

We shall discuss the chiral phase transition dynamics by including the effect of thermal fluctuations.
Thermal corrections can be calculated by using the standard imaginary time formulation of the thermal field theory
(see e.g. refs.~\cite{Quiros:1999jp,Kapusta:2006pm} for reviews).
The thermal effective potential is given by 
\begin{align}
	&V_{\rm th}^{\rm NJL}=-8N_c T \int\dfrac{\mathrm{d}^3k}{(2\pi)^3}\sum_{i=1,2}\log\left(1+\exp\left[-\dfrac{E^i_{k}}{T}\right]\right), \nonumber \\[1ex]
	&E^i_k\equiv \sqrt{k^2+\Theta(\Lambda_{\rm 3D}-k)\Lambda^2_{\rm 3D}\xi_i^2}\,,\label{eq:thermal potential NJL}
\end{align}
where $k$ denotes the magnitude of the three-dimensional momentum, and $\Theta(x)$ is the Heaviside step function.
In the above expression of the thermal potential, we introduce a 3D momentum cutoff scale,
following refs.~\cite{Kahara:2012yr,Reichert:2021cvs}.
Although the thermal potential is UV finite due to the Boltzmann suppression, this cutoff treatment may be required because we introduce it for the zero-temperature potential.\footnote{For example, in a $SU(3)$ gauge theory with fermions
in the adjoint representation, it is found in ref.~\cite{Kahara:2012yr} that this treatment is required
to obtain a clear distinction between the confinement-deconfinement and chiral phase transitions.}
However, we stress that the first-order chiral phase transition takes place for sufficiently large Fermi constants if we do not impose this cutoff treatment for the finite-temperature effective potential.
This conclusion is also found in ref.~\cite{Reichert:2021cvs}.

In the numerical analysis, it is convenient to parameterize the total effective potential,
$V_{\rm tot}^{\rm NJL}=V^{\rm NJL}_{\rm zero}+V^{\rm NJL}_{\rm th}$,
as
\begin{align}
	\bar{V}^{\rm NJL}_{\rm tot}(\bar{G}_1,\bar{G}_2,\tilde{T},\xi_1,\xi_2)
 &\equiv\dfrac{V^{\rm NJL}_{\rm tot}}{N_c\Lambda_{\rm 3D}^4}   \\[1ex]
 &= \tilde{T}^4\bar{V}_{\rm th}^{\rm NJL}+\bar{V}^{\rm NJL}_{\rm zero}(\bar{G}_1,\bar{G}_2,\xi_1,\xi_2) \, , \nonumber
\end{align}
where
\begin{align}
\bar{V}_{\rm th}^{\rm NJL}= &- \dfrac{4}{\pi^2}\int^\infty_0\mathrm{d}t\,t^2\biggl[\sum_{i=1,2}\log\biggl(1 \nonumber \\
&+\exp\left[-\sqrt{t^2+\Theta\left(1/\tilde{T}-t\right)\dfrac{\xi_i^2}{\tilde{T}^2}}\right]\biggr)\biggr] \, ,
\end{align}
with $\tilde{T}\equiv T/\Lambda_{\rm 3D}$.
We numerically evaluate $\bar{V}_{\rm tot}^{\rm NJL}$ within the range of $-50\leq\bar{G_i}\leq 50$ ($i=1,2$), and investigate the temperature dependence of potential minima.
We find that there is no parameter region that leads to the first-order phase transition.

Let us comment on the result of the current NJL analysis
by comparing that of the analysis based on the argument of universality.
In section~\ref{sec:Sp(2N)}, we have seen that the fluctuation-induced first-order phase transition is expected to take place
for the symmetry breaking pattern $U(2N)\to Sp(2N)$ for $N\geq 2$.
For $\bar{G}_2=0$, the Lagrangian of the NJL model~\eqref{eq:four-fermi interaction} possesses
the enlarged $U(4)$ symmetry, but the order of the chiral phase transition based on this model
is not of the first order, which is in tension with the result based on the universality argument.
In the NJL analysis, one takes account of the effect of thermal fluctuations of quarks on the chiral condensate
as we have explicitly performed here,
while thermal fluctuations of the chiral condensate itself have not been considered
because the mean-field approximation is assumed.
Indeed, the second-order phase transition takes place in the analysis based on the argument of universality
if we neglect fluctuations of the chiral condensate.
Therefore, in the NJL model, one may not capture the important effect originated
from fluctuations of the chiral condensate which plays a central role to determine the order of the phase transition.
It may be interesting to study the chiral phase transition by using the quark-meson model~\cite{Ellwanger:1994wy,Jungnickel:1995fp}
because thermal fluctuations of the chiral condensate, as well as fluctuations of quarks
coupled to the chiral condensate, may be adequately included by the functional RG method.

\bibliography{ref}

\end{document}